\documentclass{article} 
\usepackage[preprint]{colm2025_conference}
\usepackage[T1]{fontenc}
\usepackage{colortbl}
\usepackage{tikz}
\usepackage{pgfplots}

\usepackage{tcolorbox}
\usetikzlibrary{arrows.meta,positioning,calc,fit,decorations.pathreplacing}
\pgfplotsset{compat=1.18}

\definecolor{GroupBG}{HTML}{F1F3F6}
\definecolor{ControlBG}{HTML}{EEF3F8}
\definecolor{OursBand}{HTML}{F6E3DE}
\definecolor{OursBG}{HTML}{FCF0EC}
\definecolor{FrontierBlue}{HTML}{4C78A8}
\definecolor{FrontierOrange}{HTML}{F28E2B}
\definecolor{FrontierGreen}{HTML}{59A14F}
\definecolor{FrontierRed}{HTML}{E15759}
\definecolor{FrontierPurple}{HTML}{B07AA1}
\definecolor{BoxBG}{HTML}{F7F8FA}
\definecolor{BoxFrame}{HTML}{C7CCD4}

\newtcolorbox{rqbox}{
  colback=BoxBG,
  colframe=BoxFrame,
  boxrule=0.45pt,
  arc=1.2pt,
  left=4pt,right=4pt,top=3pt,bottom=3pt,
  before skip=4pt,after skip=4pt,
  fontupper=\small
}

\newtcolorbox{takeawaybox}{
  colback=OursBG,
  colframe=OursBand,
  boxrule=0.45pt,
  arc=1.2pt,
  left=4pt,right=4pt,top=3pt,bottom=3pt,
  before skip=4pt,after skip=4pt,
  fontupper=\small
}

\usepackage{microtype}
\usepackage{hyperref}
\usepackage{url}
\usepackage{xurl}
\usepackage{breakurl}
\usepackage{booktabs}
\usepackage{enumerate}

\usepackage{amsmath, amssymb, graphicx}
\usepackage{algorithm, algpseudocode}
\usepackage{float}

\usepackage{lineno}

\usepackage{listings}
\usepackage{xcolor}
\usepackage{multirow}
\usepackage{CJKutf8}

\tcbuselibrary{listings, skins}

\usepackage{caption}
\definecolor{darkblue}{rgb}{0, 0, 0.5}
\hypersetup{colorlinks=true, citecolor=darkblue, linkcolor=darkblue, urlcolor=darkblue}

\title{DAT: Dynamic Alpha Tuning for Hybrid Retrieval in Retrieval-Augmented Generation}

\author{%
Hsin-Ling Hsu\\
National Chengchi University\\
\texttt{112306092@nccu.edu.tw}
\And
Jerry Wang\\
University of Illinois Urbana-Champaign\\
\texttt{jerryw10@illinois.edu}
\And
Jengnan Tzeng\thanks{Corresponding author.}\\
National Chengchi University\\
\texttt{glophy@g.nccu.edu.tw}
}

\begin{document}

\ifcolmsubmission
\linenumbers
\fi

\maketitle
\lhead{The DAT Series: DAT and Frontier-DAT}

\begin{abstract}
Hybrid retrieval techniques in Retrieval-Augmented Generation (RAG) systems enhance information retrieval by combining dense and sparse (e.g., BM25-based) retrieval methods. However, existing approaches struggle with adaptability, as fixed weighting schemes fail to adjust to different queries. To address this, we propose DAT (Dynamic Alpha Tuning), a novel hybrid retrieval framework that dynamically balances dense retrieval and BM25 for each query. DAT leverages a large language model (LLM) to evaluate the effectiveness of the top-1 results from both retrieval methods, assigning an effectiveness score to each. It then calibrates the optimal weighting factor through effectiveness score normalization, ensuring a more adaptive and query-aware weighting between the two approaches. Empirical results show that DAT consistently significantly outperforms fixed-weighting hybrid retrieval methods across various evaluation metrics. Even on smaller models, DAT delivers strong performance, highlighting its efficiency and adaptability.

\end{abstract}

\section{Introduction}

Retrieval-Augmented Generation (RAG) \citep{NEURIPS2020_6b493230} systems have emerged as a powerful paradigm for enhancing the factuality and reliability of large language model (LLM) outputs by grounding responses in external knowledge sources. At the core of effective RAG systems lies the retrieval component, which is responsible for identifying and surfacing the most relevant documents from a knowledge base in response to user queries. The quality of retrieval directly impacts the overall performance of RAG systems, making it a critical area for optimization.

Hybrid retrieval \citep{Ma2020HybridFR,Sawarkar_2024,azureai2023} approaches combining sparse (e.g., BM25) and dense methods have demonstrated superior performance compared to either method alone. BM25 \citep{bm251994} excels at precise keyword matching through term frequency calculations, while dense retrieval \citep{karpukhin-etal-2020-dense} captures semantic relationships that may not involve direct lexical overlap. While the complementary strengths of these methods are well established, effectively balancing their contributions remains challenging. Current approaches \citep{Bruch_2023} typically employ a fixed weighting parameter ($\alpha$) determined through offline tuning on validation datasets. This static weighting scheme, however, fails to account for the diverse nature of user queries, where the optimal balance between keyword matching and semantic similarity varies significantly based on query characteristics and knowledge base structure.

Recent efforts to address this limitation include approaches that assign different $\alpha$ values based on query types (e.g., fact-seeking, concept-seeking, etc.) \citep{llamaindexalphatuning}. However, these methods still rely on predetermined categories with fixed weights and often overlook the complex interplay between individual queries and the knowledge base. If this assumption holds—that many queries benefit more from extreme values (i.e., pure BM25 or pure dense retrieval)—then using a compromise value such as $\alpha = 0.5$, while seemingly optimal on average, may in fact lead to suboptimal performance for most individual queries. This would pose a significant challenge to hybrid retrieval optimization.

These limitations and opportunities motivate our research questions:
\begin{itemize}
    \item \textbf{How can we effectively combine sparse and dense retrieval methods to maximize retrieval performance?}
    \item \textbf{How can a retrieval system adapt to the specific relationship between each query and the knowledge base to determine optimal retrieval parameters?}
    \item \textbf{To what extent can dynamic weight adjustment improve individual query performance compared to fixed weighting approaches?}
    \item \textbf{Can we design a hybrid retrieval system that balances effectiveness and efficiency by leveraging minimal LLM reasoning?}
\end{itemize}

\begin{figure}[htbp]
    \centering
    \includegraphics[width=\linewidth]{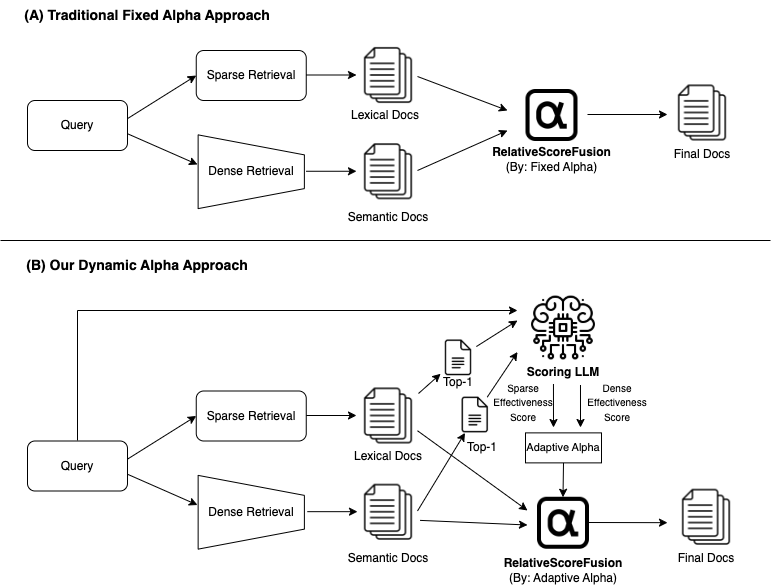}
    \caption{Comparison between the traditional fixed alpha approach (top) and our proposed DAT approach (bottom). While fixed alpha methods use predetermined weights regardless of query characteristics, DAT dynamically adjusts the retrieval weighting coefficient based on query-specific features, optimizing retrieval performance.}
    \label{fig:intro}
\end{figure}

To address these challenges, we introduce \textbf{DAT}, a novel framework that adaptively adjusts the retrieval weighting coefficient based on query-specific characteristics. The key insight of our approach is that the optimal hybrid weighting $\alpha$ for each query should reflect the relative effectiveness of sparse and dense retrieval methods in that specific context. As shown in Figure~\ref{fig:intro}, DAT dynamically computes an optimal $\alpha$ value for each query while minimizing computational overhead.

Our approach evaluates the relative effectiveness of each retrieval method by comparing the top-1 retrieved result from both BM25 and dense retrieval. This design is based on the understanding that sparse retrieval effectiveness depends primarily on term overlap between queries and documents, while dense retrieval relies on effective semantic embedding alignment. By assessing only the top result from each method with an LLM-based effectiveness scoring mechanism, we can efficiently determine which retrieval method performs better for a given query without the computational burden of evaluating multiple documents. This lightweight evaluation provides a strong signal for estimating relative retrieval strength for dynamic $\alpha$ calculation.

Our contributions are four-fold:

\begin{itemize}
    \item We propose a query-adaptive framework that dynamically calibrates the weighting between sparse and dense retrieval methods, eliminating reliance on predetermined, static $\alpha$ values.
    
    \item We introduce an LLM-based effectiveness scoring mechanism that evaluates retrieval effectiveness based on the specific relationship between each query and the knowledge base, rather than on general query characteristics.
    
    \item We demonstrate that evaluating only the top result from each retrieval method provides sufficient signal for effective weighting decisions, offering a significantly more cost-efficient approach to hybrid retrieval optimization.
    
    \item We demonstrate through extensive experiments that DAT significantly outperforms fixed-weight hybrid approaches, particularly for queries where retrieval methods exhibit varying effectiveness.
\end{itemize}

Our empirical results show that DAT consistently improves retrieval performance across various evaluation metrics, with particularly substantial gains on challenging queries where standard hybrid approaches struggle.

\section{Related Work}

Hybrid retrieval systems combine the complementary strengths of multiple retrieval methods. BM25 \citep{bm251994}, the predominant sparse retrieval algorithm, calculates relevance based on term frequency and inverse document frequency, efficiently handling exact keyword matching. Meanwhile, dense retrieval approaches \citep{karpukhin-etal-2020-dense} leverage vector embeddings to capture semantic relationships beyond lexical overlap. The effectiveness of combining sparse and dense methods has been comprehensively demonstrated in hybrid RAG systems, where improved retrieval quality significantly enhances generation accuracy \citep{Ma2020HybridFR,Sawarkar_2024,azureai2023}.

The challenge of determining optimal weighting coefficients has traditionally been addressed through offline tuning \citep{Bruch_2023}, where experiments on validation sets establish a fixed weight that is then applied universally to all future queries, regardless of their characteristics. Some approaches \citep{jeong2024adaptiveraglearningadaptretrievalaugmented} attempt refinement by classifying queries into predefined types. LlamaIndex \citep{llamaindexalphatuning} proposed assigning different weights based on predefined query categories, though this approach still relies on static weights per category.

Our work differs by leveraging LLMs' reasoning capabilities to dynamically assess retrieval quality and calibrate weighting parameters per query, without relying on predefined categories or fixed weights from offline tuning. This approach aligns with recent trends in using LLMs as judges \citep{gu2024surveyllmasajudge} but applies this concept specifically to adaptive hybrid retrieval optimization. Unlike previous approaches that optimize parameters across entire query sets, DAT addresses the limitations of fixed-weight systems by dynamically adapting to each query's unique characteristics and its relationship to the knowledge base.

\section{Hybrid Retrieval}
Hybrid retrieval combines sparse and dense retrieval to leverage both keyword matching and semantic understanding. Sparse methods like BM25 score documents based on lexical overlap:
\begin{equation} \text{BM25}(q, d) = \sum_{i=1}^{n} \text{IDF}(q_i) \cdot \frac{f(q_i, d) \cdot (k_1 + 1)}{f(q_i, d) + k_1 \cdot (1 - b + b \cdot \frac{|d|}{\text{avgdl}})} \end{equation}
Dense retrieval encodes queries and documents into vectors using embedding functions $E_q$ and $E_d$, with similarity computed via cosine similarity:
\begin{equation} \text{Sim}_{\text{dense}}(q, d) = \cos(E_q(q), E_d(d)) = \frac{E_q(q) \cdot E_d(d)}{||E_q(q)|| \cdot ||E_d(d)||} \end{equation}
To combine these methods, hybrid systems typically use techniques such as the \textbf{relativeScoreFusion algorithm} \citep{weaviatefusion} which balances the influence of different retrieval approaches. First, we normalize the similarity scores from both retrieval methods using min-max scaling:
\begin{align}
\tilde{S}_{\text{dense}}(q, d) &= \frac{\text{Sim}_{\text{dense}}(q, d) - \min_{d' \in D_{\text{dense}}(q)}(\text{Sim}_{\text{dense}}(q, d')) }{\max_{d' \in D_{\text{dense}}(q)}(\text{Sim}_{\text{dense}}(q, d')) - \min_{d' \in D_{\text{dense}}(q)}(\text{Sim}_{\text{dense}}(q, d'))} \\
\tilde{S}_{\text{BM25}}(q, d) &= \frac{\text{BM25}(q, d) - \min_{d' \in D_{\text{BM25}}(q)}(\text{BM25}(q, d')) }{\max_{d' \in D_{\text{BM25}}(q)}(\text{BM25}(q, d')) - \min_{d' \in D_{\text{BM25}}(q)}(\text{BM25}(q, d'))}
\end{align}
where $\tilde{S}_{\text{dense}}(q, d)$ and $\tilde{S}_{\text{BM25}}(q, d)$ are normalized scores in the range $[0,1]$.
Hybrid systems typically combine these normalized scores using a fixed weighting parameter $\alpha$:
\begin{equation}
R(q, d) = \alpha \cdot \tilde{S}_{\text{dense}}(q, d) + (1 - \alpha) \cdot \tilde{S}_{\text{BM25}}(q, d)
\end{equation}
This fixed $\alpha$ is determined via offline tuning on validation data and then applied uniformly to all queries. However, such a static strategy fails to account for the varying nature of user queries, limiting retrieval effectiveness across diverse scenarios.

\section{DAT}

To overcome the limitations of static weighting in hybrid retrieval, we introduce DAT, a query-adaptive framework that dynamically adjusts the retrieval weighting coefficient based on the effectiveness of each method for a given query. The central intuition is that different queries inherently favor different retrieval strategies—some require precise keyword overlap (favoring BM25), while others rely on semantic alignment (favoring dense retrieval).

DAT harnesses the reasoning capabilities of large language models (LLMs) to estimate the optimal weighting coefficient $\alpha(q)$ for each query at runtime. Unlike traditional approaches that depend on predefined heuristics or offline-tuned parameters, our method adapts on-the-fly to each query's unique interaction with the knowledge base.

Given a query $q$, we retrieve the top-1 result from both sparse and dense retrieval methods: $d_{v,1} \in D_{\text{dense}}(q)$ and $d_{b,1} \in D_{\text{BM25}}(q)$. These top-ranked documents are treated as representative indicators of each method's retrieval effectiveness for the specific query. This focused sampling strategy minimizes computational cost while providing sufficient signal to inform adaptive weighting decisions.

\subsection{LLM-Based Retrieval Effectiveness Scoring}

A key component of DAT is the use of LLMs as evaluators of retrieval quality. We posit that LLMs, with their deep semantic understanding, can assess the relevance of a retrieved document to the original query and thereby estimate each retrieval method's relative effectiveness.

To formalize this, we define a scoring function $S(q, d) = f_{\text{LLM}}(q, d)$, which returns an effectiveness score in the discrete range $\{0,1,2,3,4,5\}$—with higher values indicating greater effectiveness. The scoring rubric is carefully designed to reflect retrieval effectiveness:

\begin{itemize}
\item \textbf{5 points}: Direct hit—the retrieved document directly answers the question.
\item \textbf{3–4 points}: Good wrong result—the document is conceptually close to the correct answer, indicating high likelihood that correct answers are nearby.
\item \textbf{1–2 points}: Bad wrong result—the document is loosely related but misleading, with low likelihood that correct answers are nearby.
\item \textbf{0 points}: Completely off-track—the result is totally unrelated to the query.
\end{itemize}

Our prompting strategy (see Appendix~\ref{appendix:prompt}) guides the LLM to prioritize factual alignment and informational completeness over superficial similarity or stylistic matching. The LLM independently evaluates each of the top-1 documents and assigns scores: $S_v(q) = S(q, d_{v,1})$ for dense retrieval and $S_b(q) = S(q, d_{b,1})$ for BM25. This decoupled assessment ensures that the relative retrieval effectiveness is directly captured and can inform downstream weighting.

\subsection{Dynamic Alpha Calculation}

Using the LLM-assigned scores, we compute the dynamic weighting coefficient $\alpha(q)$ through a case-aware formulation that ensures robust behavior across various retrieval outcomes:

\begin{equation}
\alpha(q) = \begin{cases} 
0.5, & \text{if } S_v(q) = 0 \text{ and } S_b(q) = 0, \\
1.0, & \text{if } S_v(q) = 5 \text{ and } S_b(q) \neq 5, \\
0.0, & \text{if } S_b(q) = 5 \text{ and } S_v(q) \neq 5, \\
\frac{S_v(q)}{S_v(q) + S_b(q)} & \text{otherwise}.
\end{cases}
\end{equation}

This rule-based approach ensures:

\begin{itemize}
    \item Equal weighting (0.5) when both retrieval methods fail to return relevant content.
    \item Exclusive preference (1.0 or 0.0) when one method yields a perfect result and the other does not.
    \item Proportional weighting when both methods return partially relevant results.
\end{itemize}

For stability and implementation consistency, the final $\alpha(q)$ value is rounded to one decimal place before being applied in the hybrid scoring function.

\subsection{Final Score Fusion}
With the dynamically determined $\alpha(q)$, we compute the final hybrid ranking score by applying the weighted combination to the normalized scores from both retrieval methods:
\begin{equation}
R(q, d) = \alpha(q) \cdot \tilde{S}_{\text{dense}}(q, d) + (1 - \alpha(q)) \cdot \tilde{S}_{\text{BM25}}(q, d)
\end{equation}
Documents are then ranked based on $R(q, d)$, and the top-$K$ results form the final retrieval output $D_{\text{final}}(q) = \{d_1, d_2, ..., d_K\}$ that is passed to the generation component of the RAG system.
Through this dynamic approach, DAT effectively overcomes the limitations of fixed-weight hybrid retrieval methods by intelligently adapting to each query's characteristics. This query-specific adaptation leads to more relevant and accurate retrieval results across diverse query types, enhancing the overall performance of RAG systems.

\section{Experiments}

\subsection{Experimental Setup}

\paragraph{Datasets and Preprocessing}
To evaluate the effectiveness and generalizability of our proposed method, we conducted experiments on two benchmark datasets: SQuAD \citep{rajpurkar-etal-2016-squad}, a widely used dataset for evaluating retrieval-based question answering in English, and DRCD \citep{drcd}, a large-scale traditional Chinese machine reading comprehension dataset.

While SQuAD provides a standard benchmark, we include DRCD to examine whether the proposed method can also perform well in a different language setting. As Chinese is one of the most widely spoken languages, DRCD serves as a valuable testbed for assessing the broader applicability of our approach.

For each dataset, we constructed an evaluation corpus by randomly sampling articles from the original document collection. For each selected article, we included all its paragraphs $\mathcal{P} = \{p_1, p_2, \dots\}$ and the corresponding questions $\mathcal{Q} = \{q_1, q_2, \dots\}$ such that each question $q_i \in \mathcal{Q}$ is answerable by a span in paragraph $p_i \in \mathcal{P}$. The sampling process continued until the number of questions approached 3000, stopping before the next sampled article would exceed this threshold. This yields a paragraph corpus $\mathcal{P}_{\text{eval}} \subset \mathcal{P}$ and a query set $\mathcal{Q}_{\text{eval}} \subset \mathcal{Q}$, with aligned pairs $(q_i, p_i)$ forming the ground truth for retrieval.

To better focus our evaluation, we identified a subset of queries $\mathcal{Q}{\text{hybrid}} \subset \mathcal{Q}{\text{eval}}$ where hybrid retrieval strategies can actually make a difference. Our analysis revealed that for many queries in $\mathcal{Q}{\text{eval}} \setminus \mathcal{Q}{\text{hybrid}}$, retrieval performance remained identical regardless of the $\alpha$ value used—suggesting these queries were too simple to benefit from hybrid approaches and could be optimally retrieved using either BM25 or dense retrieval alone. In contrast, the $\mathcal{Q}_{\text{hybrid}}$ subset specifically contains queries where BM25 and dense retrieval produce different rankings, and where the choice of $\alpha$ directly impacts whether the correct document appears at the top position. This hybrid-sensitive subset serves as a focused testbed for evaluating the effectiveness of dynamic weighting strategies in scenarios where hybrid retrieval is truly beneficial. Detailed dataset statistics are summarized in Table~\ref{tab:dataset_statistics}.

\begin{table}[htbp]
\small
  \centering
    \begin{tabular}{lcccc}
    \toprule
        \textbf{Dataset} & \textbf{Articles} & \textbf{Paragraphs} & \textbf{Questions} & \textbf{Hybrid-Sensitive} \\
        \midrule
        SQuAD & 13 & 585 & 2976 & 1111 \\
        DRCD  & 318 & 908 & 3000 & 1523 \\
        \bottomrule
    \end{tabular}
    \caption{Dataset statistics for SQuAD and DRCD evaluation corpora.}
    \label{tab:dataset_statistics}
\end{table}

The retrieval task involves identifying the most relevant paragraph $\hat{p}_i \in \mathcal{P}_{\text{eval}}$ for each query $q_i \in \mathcal{Q}_{\text{eval}}$. A retrieval is considered successful if $\hat{p}_i$ matches the ground truth paragraph $p_i$ that contains the answer.

\paragraph{Metrics and Evaluation Protocol}
To evaluate retrieval performance, we use Precision@1 and Mean Reciprocal Rank at 20 (MRR@20). Precision@1 measures the fraction of queries where the correct answer appears as the top-ranked retrieved document. MRR@20 assesses ranking quality by computing the reciprocal rank of the first correct document (up to position 20) and averaging across queries. These metrics effectively quantify retrieval accuracy and ranking effectiveness. We conduct our evaluation in two phases: a Complete Dataset Evaluation on the entire query set $\mathcal{Q}_{\text{eval}}$ and a more focused Hybrid-Sensitive Analysis on the subset $\mathcal{Q}_{\text{hybrid}}$ where hybrid retrieval methods can make a meaningful difference.

\paragraph{Baseline Methods}
We compare our proposed DAT framework against several baseline retrieval methods. The first baseline is BM25 Only (\( \alpha = 0 \)), a sparse retrieval approach using only BM25 scores. For English (SQuAD) datasets, we use standard word tokenization, while for Chinese (DRCD) datasets, we adopt the tokenizer from \texttt{ckiplab/albert-base-chinese}\footnote{\url{https://huggingface.co/ckiplab/albert-base-chinese}}. The second baseline is Dense Only (\( \alpha = 1 \)), which ranks paragraphs based on cosine similarity between query and paragraph embeddings obtained from the \texttt{text-embedding-3-large} model \citep{text3large}. The third baseline is Fixed Hybrid ($\alpha = \alpha^\ast$), a hybrid method that linearly combines BM25 and dense scores with a fixed weighting parameter $\alpha^\ast$. For both datasets, we conducted exhaustive grid search over $\alpha$ values from 0 to 1 with a step size of 0.1, and found that $\alpha^\ast = 0.6$ maximized retrieval accuracy on both validation sets, making it the optimal fixed weighting value for our experiments.

\paragraph{Model Implementation}
For our proposed DAT method, we experiment with three different base models to demonstrate the robustness of our approach across various model sizes and architectures: GPT-4o, OpenAI's model \citep{openai2024gpt4o} estimated to have approximately 200B parameters \citep{abacha2025medecbenchmarkmedicalerror}; GPT-4o-mini, OpenAI's model \citep{openai2024gpt4omini} estimated to have approximately 8B parameters \citep{abacha2025medecbenchmarkmedicalerror}; and DeepSeek-R1-Distill-Qwen-14B, DeepSeek's 14B parameter open source model \citep{deepseekai2025deepseekr1}.

\subsection{Results}

\subsubsection{Complete Dataset Evaluation}

We first evaluate all methods on the complete datasets $\mathcal{Q}_{\text{eval}}$ for each dataset. Table~\ref{tab:complete_alpha_selection} shows the accuracy of \( \alpha \) selection for both SQuAD and DRCD, measuring how often each method selects the optimal weighting value for a given query. We define the optimal weighting value as the $\alpha$ that produces the highest retrieval ranking for the ground truth paragraph $p_i$ given query $q_i$. 

The results in Table~\ref{tab:complete_alpha_selection} show that DAT variants achieve higher alpha selection accuracy than fixed weighting approaches on both datasets, with GPT-4o reaching 0.9234 accuracy on SQuAD and GPT-4o-mini achieving 0.9013 on DRCD compared to 0.8975 and 0.8623 for the Fixed Hybrid approach, respectively.

The results in Table~\ref{tab:complete_retrieval} demonstrate that our DAT approach consistently outperforms both single-method retrieval and fixed-weight hybrid retrieval across both datasets. For SQuAD, even the DAT variant using the small model (DeepSeek-R1-Distill-Qwen-14B) achieves significant improvements over the best fixed hybrid approach, with a \textasciitilde 2\% increase in Precision@1. Similarly for DRCD, all DAT variants provide notable improvements, with GPT-4o achieving a \textasciitilde 3.3\% increase in Precision@1.

Interestingly, we observe that while dense retrieval performs relatively well on the SQuAD dataset, BM25 shows stronger performance on the DRCD dataset, suggesting different retrieval dynamics across datasets. Despite these differences, DAT effectively adapts to both scenarios by dynamically selecting appropriate alpha values.

\begin{table}[htp]
\small
  \centering
    \begin{tabular}{lccc}
    \toprule
        \textbf{Method} & \textbf{SQuAD} & \textbf{DRCD} \\
        \midrule
        BM25 Only (\( \alpha = 0.0 \)) & 0.7981 & 0.8110 \\
        Dense Only (\( \alpha = 1.0 \)) & 0.7789 & 0.6156 \\
        Fixed Hybrid (\( \alpha = 0.6 \)) & 0.8975 & 0.8623 \\
        \midrule
        DAT (DeepSeek-R1-Distill-Qwen-14B) & 0.9163 & 0.8897 \\
        DAT (GPT-4o-mini) & 0.9194 & \textbf{0.9013} \\
        DAT (GPT-4o) & \textbf{0.9234} & 0.9010 \\
        \bottomrule
    \end{tabular}
    \caption{Alpha Selection Accuracy on Complete Datasets $\mathcal{Q}_{\text{eval}}$}
    \label{tab:complete_alpha_selection}
\end{table}

\begin{table}[htp]
\small
  \centering
    \resizebox{\textwidth}{!}{
    \begin{tabular}{lcccc}
    \toprule
        \multirow{2}{*}{\textbf{Method}} & \multicolumn{2}{c}{\textbf{SQuAD}} & \multicolumn{2}{c}{\textbf{DRCD}} \\
        \cmidrule(lr){2-3} \cmidrule(lr){4-5}
        & \textbf{Precision@1} & \textbf{MRR@20} & \textbf{Precision@1} & \textbf{MRR@20} \\
        \midrule
        BM25 Only (\( \alpha = 0.0 \)) & 0.7594 & 0.8223 & 0.7630 & 0.8134 \\
        Dense Only (\( \alpha = 1.0 \)) & 0.7396 & 0.8119 & 0.5743 & 0.6708 \\
        Fixed Hybrid (\( \alpha = 0.6 \)) & 0.8461 & 0.8997 & 0.8113 & 0.8619 \\
        \midrule
        DAT (DeepSeek-R1-Distill-Qwen-14B) & 0.8663 & 0.9079 & 0.8347 & 0.8711 \\
        DAT (GPT-4o-mini) & 0.8676 & 0.9093 & 0.8417 & 0.8796 \\
        DAT (GPT-4o) & \textbf{0.8740} & \textbf{0.9130} & \textbf{0.8440} & \textbf{0.8807} \\
        \bottomrule
    \end{tabular}}
    \caption{Retrieval Performance on Complete Datasets $\mathcal{Q}_{\text{eval}}$}
    \label{tab:complete_retrieval}
\end{table}

\subsubsection{Hybrid-Sensitive Analysis}

While the complete dataset evaluation demonstrates the overall effectiveness of DAT, we now focus specifically on the hybrid-sensitive subsets $\mathcal{Q}_{\text{hybrid}}$ (1111 queries for SQuAD and 1523 for DRCD) to examine performance specifically on queries where different retrieval methods produce varying rankings, making the weighting between methods critical for optimal results.

\begin{table}[htp]
\small
  \centering
    \begin{tabular}{lccc}
    \toprule
        \textbf{Method} & \textbf{SQuAD} & \textbf{DRCD} \\
        \midrule
        BM25 Only (\( \alpha = 0.0 \)) & 0.4590 & 0.6277 \\
        Dense Only (\( \alpha = 1.0 \)) & 0.4077 & 0.2429 \\
        Fixed Hybrid (\( \alpha = 0.6 \)) & 0.7255 & 0.7288 \\
        \midrule
        DAT (DeepSeek-R1-Distill-Qwen-14B) & 0.7759 & 0.7827 \\
        DAT (GPT-4o-mini) & 0.7840 & \textbf{0.8056} \\
        DAT (GPT-4o) & \textbf{0.7948} & 0.8050 \\
        \bottomrule
    \end{tabular}
    \caption{Alpha Selection Accuracy on Hybrid-Sensitive Subsets $\mathcal{Q}_{\text{hybrid}}$}
    \label{tab:subset_alpha_selection}
\end{table}

\begin{table}[htp]
\small
  \centering
    \resizebox{\textwidth}{!}{
    \begin{tabular}{lcccc}
    \toprule
        \multirow{2}{*}{\textbf{Method}} & \multicolumn{2}{c}{\textbf{SQuAD}} & \multicolumn{2}{c}{\textbf{DRCD}} \\
        \cmidrule(lr){2-3} \cmidrule(lr){4-5}
        & \textbf{Precision@1} & \textbf{MRR@20} & \textbf{Precision@1} & \textbf{MRR@20} \\
        \midrule
        BM25 Only (\( \alpha = 0.0 \)) & 0.3906 & 0.5420 & 0.5555 & 0.6446 \\
        Dense Only (\( \alpha = 1.0 \)) & 0.3375 & 0.5143 & 0.1838 & 0.3636 \\
        Fixed Hybrid (\( \alpha = 0.6 \)) & 0.6229 & 0.7493 & 0.6507 & 0.7401 \\
        \midrule
        DAT (DeepSeek-R1-Distill-Qwen-14B) & 0.6769 & 0.7712 & 0.6967 & 0.7582 \\
        DAT (GPT-4o-mini) & 0.6805 & 0.7750 & 0.7104 & 0.7749 \\
        DAT (GPT-4o) & \textbf{0.6976} & \textbf{0.7849} & \textbf{0.7150} & \textbf{0.7771} \\
        \bottomrule
    \end{tabular}}
    \caption{Retrieval Performance on Hybrid-Sensitive Subsets $\mathcal{Q}_{\text{hybrid}}$}
    \label{tab:subset_retrieval}
\end{table}

The results in Table~\ref{tab:subset_alpha_selection} demonstrate that DAT variants achieve significantly higher alpha selection accuracy compared to fixed weighting approaches on both hybrid-sensitive subsets, with GPT-4o reaching 0.7948 accuracy for SQuAD and GPT-4o-mini achieving 0.8056 for DRCD versus approximately 0.73 for the Fixed Hybrid approach on both datasets. This improved ability to dynamically select optimal weightings directly translates to enhanced retrieval performance shown in Table~\ref{tab:subset_retrieval}.

The results on both hybrid-sensitive subsets reveal several important insights:
\begin{enumerate}
    \item The performance gap between single-method retrieval and hybrid methods is substantially wider on these subsets, highlighting the importance of effective weighting in challenging cases.
    \item DAT consistently outperforms the fixed hybrid approach, with the best variant (GPT-4o) achieving a \textasciitilde 7.5\% improvement in Precision@1 for SQuAD and a \textasciitilde 6.4\% improvement for DRCD over fixed hybrid weighting.
    \item Even the smaller model variants demonstrate significant improvements, with DeepSeek-R1-Distill-Qwen-14B achieving a \textasciitilde 5.4\% improvement in Precision@1 for SQuAD and a \textasciitilde 4.6\% improvement for DRCD over fixed hybrid weighting.
\end{enumerate}
These findings confirm that our dynamic alpha tuning approach is particularly valuable for complex queries where different retrieval methods exhibit varying effectiveness, precisely the scenarios where intelligent weighting is most needed.

\section{Scoring Mechanism Analysis}

To provide deeper insight into DAT's effectiveness, we analyze how the LLM-based scoring mechanism evaluates retrieval quality and dynamically adjusts $\alpha$ values through representative case studies from both datasets.

\subsection{SQuAD Query Example}

For the SQuAD query ``What gun did the Royal Navy start using?'', the correct answer mentions Britain's 3.7-inch HAA gun. The Top-1 result from dense retriever returned: ``By the early 20th century balloon, or airship, guns, for land and naval use were attracting attention. Various types of ammunition were proposed...'' The Top-1 result from BM25 retriever returned: ``AAA battalions were also used to help suppress ground targets. Their larger 90 mm M3 gun would prove... Also available to the Americans at the start of the war was the 120 mm M1 gun...''

The LLM's reasoning process determined:
\begin{itemize}
    \item The dense result provides topical relevance to naval guns but lacks specificity about Royal Navy adoption. Score: 3/5.
    \item The BM25 result focuses on American artillery unrelated to the Royal Navy. Score: 2/5.
\end{itemize}

This yielded $\alpha = 3/(3+2) = 0.6$, appropriately favoring the semantically-related dense result that was more likely to lead to relevant information.

\subsection{DRCD Query Example}
\begin{CJK}{UTF8}{bkai}
For the DRCD query ``水分子中的質子在高溫中與鋯進行無氧性氧化反應後什麼物質會產生?'' (What substance is produced when protons in water molecules react anaerobically with zirconium at high temperatures?), the Top-1 result from dense retriever returned: ``氫在氧化後會失去它的電子，形成氫陽離子。氫陽離子不含電子，其原子核通常只含一個質子...'' (When hydrogen is oxidized, it loses its electron and becomes a hydrogen ion...). The Top-1 result from BM25 retriever returned: ``在無氧條件下，鐵和合成鋼會被水分子中的質子緩慢氧化，而水則會還原成分子氫...'' (Under anaerobic conditions, iron and steel are oxidized by protons in water molecules, producing molecular hydrogen...).

The LLM evaluated:
\begin{itemize}
    \item The dense result explains hydrogen ions but lacks discussion of their reaction with zirconium. Score: 3/5.
    \item The BM25 result discusses a reaction mechanism with iron that parallels what occurs with zirconium, correctly identifying hydrogen gas as the product. Score: 4/5.
\end{itemize}

This produced $\alpha = 3/(3+4) = 0.43$ (rounded to 0.4), appropriately weighting toward the more relevant result.
\end{CJK}

These examples demonstrate how DAT's scoring mechanism effectively balances retrieval methods based on result quality rather than relying on predefined query categories. By evaluating the specific relationship between each query and the retrieved contents, the system adapts to diverse information needs.

\section{From Endpoint Judgments to Outcome Selection}
\label{sec:frontier-dat}

DAT judges only the top-1 results of the two endpoint retrievers ($\alpha=0$
and $\alpha=1$), yet the resulting $\alpha(q)$ can be any interior weight. An
interior weight may promote a document that neither endpoint ranks first, so
DAT can execute a ranking it has never observed. Appendix~\ref{app:frontier-dat}
formalizes this observation--action mismatch and introduces
\textbf{Frontier-DAT}. For a fixed candidate pool, linear fusion induces only
finitely many distinct rankings as $\alpha$ varies; Frontier-DAT enumerates
these executable outcomes, lets the LLM judge select one, and maps it back to a
feasible weight. On four additional benchmarks (NQ, FiQA, HotpotQA, and
MuSiQue), selecting over executable outcomes raises NQ Hit@1 over DAT from
.416 to .438, and the gain persists under matched confidence gating (.426 vs.\ .443). With top-5 Frontiers, all-support@5 improves over endpoint-only slates by
7.0 points on HotpotQA and 5.5 points on MuSiQue, suggesting that interior
weights matter most when retrieval requires complementary evidence.

\section{Conclusion}

We introduced DAT, a framework that dynamically adjusts the weighting between sparse and dense retrieval for each query by leveraging LLMs to evaluate document effectiveness. Unlike static weighting schemes, DAT adaptively selects the optimal $\alpha$ per query, achieving a balance between performance and efficiency—even with smaller models. Experiments show that DAT consistently outperforms fixed-weight hybrids, especially on hybrid-sensitive queries, and remains robust across different LLM sizes. These results underscore the limitations of static approaches and highlight the value of query-adaptive strategies. Appendix~\ref{app:frontier-dat} further extends DAT from weight prediction to selection over executable retrieval outcomes.

\section*{Author Contributions}
Hsin-Ling Hsu developed DAT and, together with Jerry Wang, the Frontier-DAT extension. Jerry Wang contributed to the Frontier-DAT formulation and experiments. Jengnan Tzeng supervised the project.

\bibliography{colm2025_conference}
\bibliographystyle{colm2025_conference}

\appendix

\section{DAT Prompt Template}\label{appendix:prompt}

\begin{lstlisting}
You are an evaluator assessing the retrieval effectiveness of dense retrieval (Cosine Distance) and BM25 retrieval for finding the correct answer.

## Task:
Given a question and two top1 search results (one from dense retrieval, one from BM25 retrieval), score each retrieval method from **0 to 5** based on whether the correct answer is likely to appear in top2, top3, etc.

### **Scoring Criteria:**
1. **Direct hit --> 5 points**  
   - If the retrieved document directly answers the question, assign **5 points**.  

2. **Good wrong result (High likelihood correct answer is nearby) --> 3-4 points**  
   - If the top1 result is **conceptually close** to the correct answer (e.g., mentions relevant entities, related events, partial answer), it indicates the search method is in the right direction.  
   - Give **4** if it's very close, **3** if somewhat close.

3. **Bad wrong result (Low likelihood correct answer is nearby) --> 1-2 points**  
   - If the top1 result is **loosely related but misleading** (e.g., shares keywords but changes context), correct answers might not be in top2, top3.  
   - Give **2** if there's a small chance correct answers are nearby, **1** if unlikely.

4. **Completely off-track --> 0 points**  
   - If the result is **totally unrelated**, it means the retrieval method is failing.  

---

### **Given Data:**
- **Question:** "{question}"
- **dense retrieval Top1 Result:** "{vector_reference}"
- **BM25 retrieval Top1 Result:** "{bm25_reference}"

---

### **Output Format:**  
Return two integers separated by a space:  
- **First number:** dense retrieval score.  
- **Second number:** BM25 retrieval score.  
- Example output: 3 4  
  (Vector: 3, BM25: 4)  

**Do not output any other text.**
\end{lstlisting}

\section{Frontier-DAT}
\label{app:frontier-dat}
\suppressfloats[t]

\begingroup
\emergencystretch=3em

\subsection{Problem Formulation}

For each query $q$, let $\mathcal C(q)$ be the finite candidate pool produced by
a fixed retrieval substrate. In our experiments it is the union of sparse and
dense top-$100$ results; all exactness statements below are conditional on
$\mathcal C(q)$ and its stored sparse/dense scores. Let
$\tilde s_s(q,d)$ and $\tilde s_d(q,d)$ denote normalized scores for
$d\in\mathcal C(q)$. Linear hybrid retrieval ranks candidates by
\begin{equation}
 s(q,d;\alpha)=(1-\alpha)\tilde s_s(q,d)+\alpha\tilde s_d(q,d),
 \qquad \alpha\in[0,1].
 \label{eq:fdapp-hybrid}
\end{equation}
Let $S_k(q;\alpha)$ be the ordered top-$k$ ranking induced by
Eq.~\eqref{eq:fdapp-hybrid}. A fixed hybrid retriever executes one tuned
$\alpha_g$.

\begin{figure}[htbp]
\centering
\begin{minipage}[t]{0.535\textwidth}
\centering
\begin{tikzpicture}
\begin{axis}[
    width=\linewidth,
    height=4.15cm,
    xmin=0,xmax=1,
    ymin=0.15,ymax=1.04,
    xlabel={fusion weight $\alpha$},
    ylabel={hybrid score},
    xtick={0,0.28,0.47,0.66,1},
    xticklabels={$0$,$\beta_1$,$\beta_2$,$\beta_3$,$1$},
    ytick=\empty,
    axis line style={draw=black!55},
    tick style={draw=black!55},
    label style={font=\scriptsize},
    tick label style={font=\scriptsize},
    clip=false,
]
  \path[fill=FrontierBlue!7]
      (axis cs:0,0.15) rectangle (axis cs:0.28,1.04);
  \path[fill=FrontierOrange!8]
      (axis cs:0.28,0.15) rectangle (axis cs:0.47,1.04);
  \path[fill=FrontierGreen!8]
      (axis cs:0.47,0.15) rectangle (axis cs:0.66,1.04);
  \path[fill=FrontierPurple!8]
      (axis cs:0.66,0.15) rectangle (axis cs:1,1.04);

  \addplot[black!25,thin,domain=0:1,samples=2]
      {0.96-0.75*x};
  \addplot[black!25,thin,domain=0:1,samples=2]
      {0.82-0.25*x};
  \addplot[black!25,thin,domain=0:1,samples=2]
      {0.68+0.05*x};
  \addplot[black!25,thin,domain=0:1,samples=2]
      {0.35+0.55*x};

  \addplot[FrontierBlue,very thick,domain=0:0.28,samples=2]
      {0.96-0.75*x};
  \addplot[FrontierOrange,very thick,domain=0.28:0.47,samples=2]
      {0.82-0.25*x};
  \addplot[FrontierGreen,very thick,domain=0.47:0.66,samples=2]
      {0.68+0.05*x};
  \addplot[FrontierPurple,very thick,domain=0.66:1,samples=2]
      {0.35+0.55*x};

  \draw[densely dashed,black!30]
      (axis cs:0.28,0.15)--(axis cs:0.28,1.02);
  \draw[densely dashed,black!30]
      (axis cs:0.47,0.15)--(axis cs:0.47,1.02);
  \draw[densely dashed,black!30]
      (axis cs:0.66,0.15)--(axis cs:0.66,1.02);

  \node[font=\scriptsize\bfseries,text=FrontierBlue]
      at (axis cs:0.09,0.98) {A};
  \node[font=\scriptsize\bfseries,text=FrontierOrange]
      at (axis cs:0.36,0.765) {C};
  \node[font=\scriptsize\bfseries,text=FrontierGreen]
      at (axis cs:0.555,0.742) {D};
  \node[font=\scriptsize\bfseries,text=FrontierPurple]
      at (axis cs:0.87,0.88) {B};

  \addplot[only marks,mark=*,mark size=1.8pt,black]
      coordinates {(0,0.96) (1,0.90)};

  \node[font=\tiny,anchor=south west]
      at (axis cs:0.01,1.005) {$S_1(0)$};
  \node[font=\tiny,anchor=south east]
      at (axis cs:0.99,0.925) {$S_1(1)$};

  \draw[densely dashed,FrontierRed,thick]
      (axis cs:0.56,0.15)--(axis cs:0.56,0.715);

  \node[
      font=\tiny\bfseries,
      align=center,
      text=FrontierRed!90!black,
      fill=FrontierRed!12,
      draw=FrontierRed!45,
      rounded corners=2pt,
      line width=0.3pt,
      inner xsep=3.5pt,
      inner ysep=2pt,
      anchor=west
  ] at (axis cs:0.60,0.34)
      {interior $\hat\alpha$\\executes D};

  \node[font=\tiny,anchor=north]
      at (axis cs:0.14,0.18) {A};
  \node[font=\tiny,anchor=north]
      at (axis cs:0.375,0.18) {C};
  \node[font=\tiny,anchor=north]
      at (axis cs:0.565,0.18) {D};
  \node[font=\tiny,anchor=north]
      at (axis cs:0.83,0.18) {B};
\end{axis}
\end{tikzpicture}
\end{minipage}
\hfill
\begin{minipage}[t]{0.435\textwidth}
\centering
\resizebox{\linewidth}{!}{%
\begin{tikzpicture}[
  >=Latex,
  font=\scriptsize,
  box/.style={
    draw,
    rounded corners=2pt,
    minimum height=6mm,
    inner xsep=4pt,
    align=center
  },
  arrow/.style={->,thick}
]
  \node[font=\scriptsize\bfseries,anchor=west]
      at (0,3.25) {Endpoint view};

  \node[box,fill=ControlBG,minimum width=1.45cm]
      (eobs) at (0.85,2.60) {observe\\$\{A,B\}$};
  \node[box,minimum width=1.10cm]
      (ej) at (2.35,2.60) {judge};
  \node[box,minimum width=1.05cm]
      (ea) at (3.70,2.60) {$\hat\alpha$};
  \node[
      box,
      draw=FrontierRed,
      densely dashed,
      text=FrontierRed,
      minimum width=1.30cm
  ] (ed) at (5.25,2.60) {execute\\D};

  \draw[arrow] (eobs)--(ej);
  \draw[arrow] (ej)--(ea);
  \draw[arrow] (ea)--(ed);

  \node[
      font=\tiny,
      text=FrontierRed,
      anchor=north,
      align=center
  ] at (5.25,2.22)
      {outcome first appears\\after the decision};

  \node[font=\scriptsize\bfseries,anchor=west]
      at (0,1.45) {Outcome-space view};

  \node[box,fill=OursBG,minimum width=1.75cm]
      (fobs) at (0.95,0.80) {enumerate\\$\{A,C,D,B\}$};
  \node[box,fill=OursBG,minimum width=1.35cm]
      (fj) at (2.80,0.80) {judge\\selects D};
  \node[box,fill=OursBG,minimum width=1.15cm]
      (fi) at (4.35,0.80) {$I_D$};
  \node[box,fill=OursBG,minimum width=1.35cm]
      (fa) at (5.95,0.80) {execute\\$\hat\alpha\in I_D$};

  \draw[arrow] (fobs)--(fj);
  \draw[arrow] (fj)--(fi);
  \draw[arrow] (fi)--(fa);

  \node[font=\tiny,anchor=north,align=center]
      at (3.45,0.36)
      {inspect outcome first; then choose a feasible weight};
\end{tikzpicture}%
}
\end{minipage}
\caption{\textbf{From endpoint judgments to outcome-space control.}
\textbf{Left:} affine hybrid scores induce a sequence of executable winners;
C and D occur only at interior weights even though an endpoint-conditioned
controller observes only A and B. \textbf{Right:} endpoint adaptation chooses a
weight before seeing its induced outcome, whereas Frontier-DAT enumerates the
executable outcomes first, selects one, and then maps it to a feasible interval
of weights.}
\label{fig:fdapp-frontier}
\end{figure}

\subsection{Top-1: Convexity Characterizes Every Executable Winner}

For a fixed query, represent each candidate by its two retrieval scores and
each fusion weight by a nonnegative direction,
\begin{equation}
 z_d=(\tilde s_s(q,d),\tilde s_d(q,d)),\qquad
 w_\alpha=(1-\alpha,\alpha),\qquad
 s(q,d;\alpha)=w_\alpha^\top z_d.
 \label{eq:fdapp-scorepoint}
\end{equation}
Top-1 fusion is therefore a two-dimensional linear optimization problem over
the score points, a classical convex-hull setting
\citep{10.1145/342009.335433}.

\paragraph{Proposition 1 (supported-hull characterization).}
Let $H_q=\operatorname{conv}\{z_d:d\in\mathcal C(q)\}$. Ignoring exact ties, a
document is top-1 for some nonzero interval of $\alpha\in[0,1]$ iff its score
point is a supported vertex of $H_q$ whose normal cone intersects
$W=\{(1-\alpha,\alpha):\alpha\in[0,1]\}$. Exact ties correspond to exposed
faces and are resolved by the retrieval substrate's deterministic tie rule.

The result follows because maximizing $w_\alpha^\top z$ over a finite set is
equivalent to maximizing it over its convex hull, where a linear functional is
optimized on an exposed face. In the equivalent slope--intercept view, the
same winners are precisely the score lines on the upper envelope. Thus the
convex geometry does not approximate the top-1 action space: conditional on
$\mathcal C(q)$, it enumerates every distinct winner that any continuous
mixture can execute.

The feasible weights for winner $d$ are
\begin{equation}
 I_d=\bigcap_{j\neq d}
 \{\alpha\in[0,1]:w_\alpha^\top(z_d-z_j)\ge0\}.
 \label{eq:fdapp-top1interval}
\end{equation}
Each constraint is a one-dimensional half-interval, so $I_d$ is an interval.
Consequently a winner cannot disappear and later reappear as $\alpha$ moves
from 0 to 1. This contiguity turns the apparent continuum of weights into an
ordered sequence of executable top-1 outcomes.

\subsection{Top-$k$: The Same Linear Structure Induces a Finite Ranking Arrangement}

For $k>1$, convex-hull vertices alone are insufficient because documents below
the upper envelope can still enter the returned slate. The linear structure,
however, still controls every ranking change. For any pair $i,j$,
\begin{equation}
 g_{ij}(\alpha)=s(q,d_i;\alpha)-s(q,d_j;\alpha)
 \label{eq:fdapp-pairdiff}
\end{equation}
is affine, so their relative order can flip at most once. Let $\mathcal B_q$
contain all pairwise crossing weights in $[0,1]$. Between consecutive
breakpoints, the sign of every $g_{ij}$ is fixed; therefore the entire ranking,
and hence every prefix $S_k(q;\alpha)$, is fixed.

\paragraph{Proposition 2 (finite ordered top-$k$ frontier).}
For $n=|\mathcal C(q)|$ pooled candidates and ignoring exact ties,
$\mathcal B_q$ partitions $[0,1]$ into at most $\binom{n}{2}+1$ cells, with a
constant ordered top-$k$ ranking inside each cell. For a fixed ordered slate
$o=(d_1,\ldots,d_k)$,
\begin{equation}
 I(o)=\{\alpha:
 s_{d_r}(\alpha)\ge s_{d_{r+1}}(\alpha)\ \forall r<k,\ 
 s_{d_k}(\alpha)\ge s_j(\alpha)\ \forall j\notin o\}
 \label{eq:fdapp-slateinterval}
\end{equation}
is an interval because it is an intersection of affine half-intervals. Thus an
ordered slate also cannot disappear and later reappear.

\subsection{The Frontier as a Canonical Decision Space}

We define the \emph{Actionable Ranking Frontier}
\begin{equation}
\begin{aligned}
\mathcal F_k(q)&=\{S_k(q;\alpha):\alpha\in[0,1]\},\\
\alpha\sim_k\alpha'&\iff S_k(q;\alpha)=S_k(q;\alpha').
\end{aligned}
\label{eq:fdapp-frontier}
\end{equation}
Conditional on the fixed candidate pool, $\mathcal F_k(q)$ is the exact quotient
$[0,1]/\!\sim_k$: it merges weights only when they have the same top-$k$
retrieval consequence. This gives a useful sufficiency property. For any
retrieval utility $U(q,S_k)$ that depends only on the returned ordered top-$k$
list,
\begin{equation}
 \max_{\alpha\in[0,1]} U(q,S_k(q;\alpha))
 =\max_{o\in\mathcal F_k(q)} U(q,o).
 \label{eq:fdapp-sufficiency}
\end{equation}
Thus optimizing over the continuous weight space loses no top-$k$ decision
information when replaced by optimization over the Frontier. The geometry is
therefore the exact realization of an outcome-space formulation, not a claim
that computational geometry itself is new.

The quotient also clarifies frontier depth. Equality under $\sim_{k+1}$
implies equality under $\sim_k$, so the partition at depth $k+1$ refines the
one at depth $k$. Increasing $k$ can reveal distinctions between previously
equivalent actions, but cannot erase an existing distinction. Whether that
finer action space is useful depends on the downstream retrieval objective.

\subsection{Frontier-DAT Procedure}

Frontier-DAT replaces direct weight prediction with selection over executable
retrieval outcomes. Given a query $q$, we first construct the Frontier
$\mathcal{F}_k(q)$ by computing all score-ordering changes over
$\alpha\in[0,1]$, evaluating one representative point from each induced cell,
and deduplicating identical ordered top-$k$ outcomes.

The resulting outcomes are then presented to a judge with the underlying
retriever identity and fusion weight hidden. For top-1, the judge selects among
the unique executable documents. For top-$k$, each unique document is shown
once with an anonymous identifier, followed by the executable ordered slates;
the judge selects the slate expected to provide the most useful collective
evidence for the query.

Each selected outcome $o\in\mathcal{F}_k(q)$ corresponds to a feasible interval
$I(o)\subseteq[0,1]$. Frontier-DAT maps the selected outcome back to an
executable fusion action by choosing the feasible weight closest to the global
fusion policy $\alpha_g$:
\begin{equation}
\hat{\alpha}=\operatorname{proj}_{I(o)}(\alpha_g).
\label{eq:fdapp-frontier-projection}
\end{equation}
Thus, Frontier construction determines which outcomes are executable, the
judge selects among those outcomes, and the selected outcome is finally mapped
back to a valid fusion weight. If the Frontier contains only one outcome, no
judgment is required.

\subsection{Experimental Setup}

We evaluate the NQ~\citep{kwiatkowski-etal-2019-natural},
FiQA~\citep{10.1145/3184558.3192301},
HotpotQA~\citep{yang-etal-2018-hotpotqa}, and
MuSiQue~\citep{trivedi-etal-2022-musique} sets, with NQ and FiQA following the
BEIR retrieval setup~\citep{thakur2021beirheterogenousbenchmarkzeroshot}.
All methods use the same fixed candidate pool, formed by the union of
BM25~\citep{10.1561/1500000019} top-100 and BGE
top-100~\citep{xiao2024cpackpackedresourcesgeneral}, with frozen
min--max-normalized scores. The tuned global weights are
$\alpha_g=1.0/\allowbreak{}0.8/\allowbreak{}0.5/\allowbreak{}0.8$
for NQ/\allowbreak{}FiQA/\allowbreak{}HotpotQA/\allowbreak{}MuSiQue.
Frontier-DAT uses GPT-4.1-mini~\citep{openai2025gpt41} at temperature 0 with
blinded retriever and $\alpha$ identities.

We compare against sparse, dense, global fusion,
RRF~\citep{10.1145/1571941.1572114},
faithful DAT (Section~4), and
QDAP-S~\citep{make8040091}. We additionally report a
continuous-$\alpha$ oracle and an Endpoint-Slate control for the top-$k$
setting. For top-1, we report Hit@1. For top-$k$, NQ and FiQA use Recall@5,
which measures whether relevant evidence is retrieved within the top five,
while HotpotQA and MuSiQue use all-support@5, which requires all annotated
supporting documents for a multi-hop question to appear in the top five.

\begin{table}[htbp]
\centering
\scriptsize
\renewcommand{\arraystretch}{0.92}
\setlength{\tabcolsep}{2.6pt}
\begin{tabular*}{\textwidth}{
  l @{\extracolsep{\fill}} c c c c c c c c
}
\toprule
\multirow{2}{*}{\textbf{Method}}
& \multicolumn{4}{c}{\textbf{Top-1 Hit@1}}
& \multicolumn{4}{c}{\textbf{Top-5 task metric}}\\
\cmidrule(lr){2-5}\cmidrule(lr){6-9}
& \textbf{NQ} & \textbf{FiQA} & \textbf{Hotpot} & \textbf{MuSiQue}
& \textbf{NQ} & \textbf{FiQA} & \textbf{Hotpot} & \textbf{MuSiQue}\\
\midrule
\rowcolor{GroupBG}
\multicolumn{9}{c}{\itshape Fixed and generic fusion baselines}\\
Sparse & .180 & .242 & .768 & .582 & .380 & .258 & .353 & .164\\
Dense & .390 & .410 & .840 & .673 & .649 & .414 & .533 & .182\\
RRF & .310 & .352 & .863 & .691 & .572 & .362 & .515 & .245\\
Global $\alpha$
& .390 & .428 & .898 & \textbf{.736}
& .649 & .424 & .567 & .227\\
\rowcolor{GroupBG}
\multicolumn{9}{c}{\itshape Adaptive and controlled baselines}\\
DAT (faithful)
& .416 & .416 & .903 & \textbf{.736}
& .631 & .414 & .557 & .255\\
QDAP-S (BGE)
& .330 & .364 & .898 & .718
& .597 & .390 & .567 & .273\\
Endpoint-Slate
& .384 & .406 & .857 & \textbf{.736}
& .670 & .417 & .552 & .236\\
\rowcolor{OursBand}
\multicolumn{9}{c}{\itshape Ours}\\
\rowcolor{OursBG}
\textbf{Top-1 Frontier-DAT}
& \textbf{.443} & \textbf{.454} & \textbf{.912} & \textbf{.736}
& .650 & \textbf{.425} & .572 & .227\\
\rowcolor{OursBG}
\textbf{Top-5 Frontier-DAT}
& .407 & .396 & .895 & .718
& \textbf{.688} & .424 & \textbf{.622} & \textbf{.291}\\
\midrule
Oracle (Upper Bound)
& .506 & .510 & .942 & .845
& .754 & .492 & .640 & .318\\
\bottomrule
\end{tabular*}
\caption{\textbf{Main results.}
Left: top-1 Hit@1. Right: the task-aligned top-5 metric (Recall@5 for NQ/FiQA;
all-support@5 for HotpotQA/MuSiQue). Bold marks the best non-oracle result in
each column. Top-1 Frontier reports the final gated policy with 2-fold cross-fitting;
Table~\ref{tab:fdapp-gatefair} separately isolates proposal quality from gating.}
\label{tab:fdapp-main}
\end{table}

\subsection{RQ1: When Does a Deeper Frontier Help?}

Table~\ref{tab:fdapp-main} compares Endpoint-Slate with Top-5 Frontier-DAT under the
same listwise judging setup, changing only whether the judge can select from
the two endpoint rankings or from the full set of executable top-5 rankings.
The deeper Frontier matters most on the multi-hop tasks: relative to
Endpoint-Slate, all-support@5 improves by 7.0 points on HotpotQA and 5.5 points
on MuSiQue, while the gains on NQ and FiQA are much smaller.

This pattern follows the retrieval objective. HotpotQA and MuSiQue reward joint
coverage of multiple supporting documents, so interior fusion weights can
create useful evidence configurations that neither endpoint ranking exposes.
Frontier depth is therefore best viewed as decision resolution, with larger
$k$ helping when the downstream utility distinguishes among richer evidence sets.

\begin{rqbox}
\textbf{RQ1 takeaway.}
Deeper Frontiers matter most when retrieval utility depends on complementary
evidence configurations.
\end{rqbox}

\subsection{RQ2: Does Outcome-Space Observation Improve the Proposal?}

RQ2 asks whether the top-1 advantage in Table~\ref{tab:fdapp-main} comes from
observing executable outcomes, rather than from greater candidate exposure or
a stronger confidence fallback.

\paragraph{Outcome observability versus candidate exposure.}
Table~\ref{tab:fdapp-obscontrol} holds the ungated judging protocol fixed while
changing what the judge can observe. Endpoint winners expose only the two
endpoint outcomes; the exposure-matched union shows more documents without
requiring them to correspond to executable fusion outcomes; Exact Frontier
shows the outcomes induced by valid fusion actions.

\begin{table}[t]
\centering
\scriptsize
\renewcommand{\arraystretch}{0.94}
\setlength{\tabcolsep}{3.0pt}
\begin{tabular*}{\columnwidth}{l@{\extracolsep{\fill}}ccc}
\toprule
\textbf{Representation} & \textbf{NQ} & \textbf{FiQA} & \textbf{Hotpot}\\
\midrule
Endpoint winners & .412 & .430 & .903\\
Exposure-matched union & .428 & .438 & .905\\
\rowcolor{OursBG}
\textbf{Exact Frontier} & \textbf{.438} & \textbf{.448} & \textbf{.912}\\
\bottomrule
\end{tabular*}
\caption{\textbf{Top-1 observation-space controls (Hit@1, ungated).}
The executable Frontier is ahead of the exposure-matched union on all three
datasets.}
\label{tab:fdapp-obscontrol}
\end{table}

As Table~\ref{tab:fdapp-obscontrol} shows, exposing more candidates already
recovers part of the gain over endpoint winners, but the executable Frontier
remains ahead of the exposure-matched union on all three datasets (+.010,
+.010, and +.007). These margins are small and not individually significant,
so we read them as consistent evidence that executability adds to exposure
rather than as a decisive separation.

\paragraph{Proposal quality versus trust calibration.}
Table~\ref{tab:fdapp-gatefair} then applies the same verifier-margin gate to DAT and
Frontier, separating proposal quality from the effect of fallback calibration.
The table shows that Frontier already produces stronger ungated proposals on
NQ and FiQA. After matched calibration, the NQ advantage persists, whereas
the FiQA gap largely closes because DAT benefits more from the gate.
Outcome observation and trust calibration therefore improve different parts
of the decision process.

\begin{table}[t]
\centering
\scriptsize
\renewcommand{\arraystretch}{0.96}
\setlength{\tabcolsep}{2.7pt}
\begin{tabular*}{\columnwidth}{l@{\extracolsep{\fill}}cccccc}
\toprule
& \multicolumn{2}{c}{\textbf{NQ}}
& \multicolumn{2}{c}{\textbf{FiQA}}
& \multicolumn{2}{c}{\textbf{Hotpot}}\\
\cmidrule(lr){2-3}\cmidrule(lr){4-5}\cmidrule(lr){6-7}
\textbf{Method}
& \textbf{Ung.} & \textbf{Gate}
& \textbf{Ung.} & \textbf{Gate}
& \textbf{Ung.} & \textbf{Gate}\\
\midrule
DAT & .416 & .426 & .416 & .446 & .903 & .907\\
\rowcolor{OursBG}
Frontier & .438 & .443 & .448 & .454 & .912 & .912\\
\midrule
Frontier$-$DAT
& \textbf{+.022} & \textbf{+.017}
& \textbf{+.032} & +.008
& +.008 & +.005\\
\bottomrule
\end{tabular*}
\caption{\textbf{Matched-gate top-1 comparison (Hit@1).}
Both methods use the same verifier gate. Bold differences have paired 95\%
confidence intervals excluding zero.}
\label{tab:fdapp-gatefair}
\end{table}

\begin{rqbox}
\textbf{RQ2 takeaway.}
Outcome-space observation gives a consistent, if modest, edge beyond candidate
exposure and, on NQ, remains beneficial under matched trust calibration.
\end{rqbox}

\subsection{Where Does the Frontier Gain Come From?}

Table~\ref{tab:fdapp-mechanism} connects the empirical gains to the geometry of the
fusion space. Interior-only winners are common, but only a subset corrects a
global-policy error. Those useful cases nevertheless account for a substantial
share of the available oracle headroom, showing that the value of the Frontier
comes from exposing rare but consequential interior outcomes rather than from
enumerating alternatives for their own sake.

\begin{table}[t]
\centering
\scriptsize
\renewcommand{\arraystretch}{0.90}
\setlength{\tabcolsep}{3.0pt}
\begin{tabular*}{\columnwidth}{l@{\extracolsep{\fill}}ccc}
\toprule
\textbf{Diagnostic} & \textbf{NQ} & \textbf{FiQA} & \textbf{Hotpot}\\
\midrule
Interior-only winner & 45\% & 44\% & 12\%\\
Useful when global wrong & 6.4\% & 3.0\% & 0.8\%\\
Oracle gap share & 55\% & 37\% & 19\%\\
\bottomrule
\end{tabular*}
\caption{\textbf{Top-1 geometric diagnostics.} ``Gap share'' is the fraction
of candidate-pool continuous-$\alpha$ oracle headroom attributable to useful
interior-only winners.}
\label{tab:fdapp-mechanism}
\end{table}

\subsection{How Strong Must the Judge Be?}

\begin{figure}[htbp]
\centering
\begin{tikzpicture}
\begin{axis}[
  width=\columnwidth,
  height=3.35cm,
  ybar,
  bar width=3.0pt,
  ymin=-30,ymax=55,
  ylabel={Oracle-gap recovery (\%)},
  symbolic x coords={L3B,Q8B,Q14B,Q32B,G4.1m},
  xtick=data,
  x tick label style={font=\scriptsize},
  y tick label style={font=\scriptsize},
  label style={font=\scriptsize},
  legend style={
    font=\scriptsize,
    at={(0.5,1.02)},
    anchor=south,
    legend columns=3,
    draw=none,
    /tikz/every even column/.append style={column sep=3pt}
  },
  axis x line*=bottom,
  axis y line*=left,
  grid=major,
  major grid style={dotted,black!20}
]
\addplot[fill=FrontierBlue,draw=FrontierBlue]
  coordinates {(L3B,22)(Q8B,42)(Q14B,42)(Q32B,49)(G4.1m,46)};
\addplot[fill=FrontierOrange,draw=FrontierOrange]
  coordinates {(L3B,-22)(Q8B,24)(Q14B,17)(Q32B,37)(G4.1m,32)};
\addplot[fill=FrontierGreen,draw=FrontierGreen]
  coordinates {(L3B,-8)(Q8B,19)(Q14B,19)(Q32B,27)(G4.1m,31)};
\coordinate (zeroL) at (axis cs:L3B,0);
\coordinate (zeroR) at (axis cs:G4.1m,0);
\draw[black!70,line width=0.8pt]
  ([xshift=-15pt]zeroL)--([xshift=15pt]zeroR);
\legend{NQ,FiQA,HotpotQA}
\end{axis}
\end{tikzpicture}
\caption{\textbf{Judge capability controls Frontier utility.}
Performance moves from an insufficient 3B judge to a positive 8B-and-above
regime, without a monotonic relation to parameter count.}
\label{fig:fdapp-judge}
\end{figure}

Figure~\ref{fig:fdapp-judge} holds the Frontier pipeline fixed and changes only the
judge. All models are evaluated in non-thinking mode, so the comparison reflects
their direct ranking ability rather than additional test-time reasoning.
Llama-3.2-3B~\citep{meta2024llama32} can turn the richer action set into harm,
whereas every tested Qwen3 model from 8B to 32B~\citep{yang2025qwen3technicalreport}
and GPT-4.1-mini operates in a positive regime across all three datasets.
In our setting, this places the empirical usability threshold roughly around
the 8B scale: below it, the judge is not consistently reliable enough to exploit
the finer distinctions exposed by the Frontier, while above it additional model
size does not monotonically improve recovery. This suggests that Frontier
utility depends primarily on reaching sufficient ranking judgment quality,
rather than on scaling the judge indefinitely.

\subsection{Experience-Adaptive Trust}

A question is whether Frontier-DAT can improve not only its retrieval
decisions, but also how much it trusts those decisions over time.
This introduces a limited self-improvement loop. Retrieval, Frontier
construction, and the judge remain frozen, while experience updates whether a
proposed override should be accepted or rejected. We model this trust decision
as a two-action contextual bandit that observes only the reward of the action
actually executed.

Table~\ref{tab:fdapp-bandit} shows that contextual
LinUCB~\citep{10.1145/1772690.1772758}
consistently improves over context-free UCB, with decisions becoming stronger
later in the stream. The frozen offline gate remains better cumulatively,
highlighting an important distinction for self-improving retrieval systems.
Adapting from experience does not automatically produce better optimization,
because online improvement must first repay cold-start and exploration costs.

\begin{table}[htbp]
\centering
\scriptsize
\renewcommand{\arraystretch}{0.92}
\setlength{\tabcolsep}{3.0pt}
\begin{tabular*}{\columnwidth}{l@{\extracolsep{\fill}}ccc}
\toprule
\textbf{Policy} & \textbf{NQ} & \textbf{FiQA} & \textbf{Hotpot}\\
\midrule
Context-free UCB & .424 & .440 & .906\\
\rowcolor{OursBG}
\textbf{Contextual LinUCB}
& .430 & .448 & .911\\
Frozen offline gate & .443 & .454 & .912\\
\midrule
Late$-$early improvement & +.014 & +.011 & +.012\\
\bottomrule
\end{tabular*}
\caption{\textbf{Experience-adaptive trust.} Cumulative Hit@1 over 30 streams.}
\label{tab:fdapp-bandit}
\end{table}

\begin{takeawaybox}
\textbf{Takeaway.}
Experience can improve future trust decisions, but online adaptation need not
beat a strong offline policy cumulatively when feedback is partial.
\end{takeawaybox}

\endgroup

\section{Limitations}
\label{app:limitations}

\paragraph{DAT.} DAT's LLM-based scoring adds latency and cost over
fixed-weight fusion. More fundamentally, it observes only the endpoint results
but may execute an interior weight whose ranking it has never seen, a gap that
Frontier-DAT (Appendix~\ref{app:frontier-dat}) addresses.

\paragraph{Frontier-DAT.} The Frontier is exact only for a fixed candidate pool
and linear score fusion, and its benefit depends on judge quality. Our evaluation uses four English benchmarks
and one retriever pair; other retrievers and end-to-end RAG quality
remain to be tested.

\end{document}